\documentclass[showpacs,aps,pre]{revtex4}

\usepackage{graphics}
\usepackage{amsmath}
\usepackage{graphicx,color}

\begin{document}

\title{A Landscape Analysis of Constraint Satisfaction Problems}

\vskip .27in

\author {Florent Krzakala$^1$ and Jorge Kurchan$^2$}
\affiliation{
PCT $^1$, CNRS UMR Gulliver 7083 and PMMH $^2$, CNRS UMR 7636 \\ 
ESPCI, 10 rue Vauquelin,
 75005 Paris, FRANCE}
\vskip .2in

\date\today

%\narrowtext

\begin{abstract}
  We discuss an analysis of Constraint Satisfaction problems, such as Sphere
  Packing, K-SAT and Graph Coloring, in terms of an effective energy
  landscape.  Several intriguing geometrical properties of the solution space
  become in this light familiar in terms of the well-studied ones of rugged
  (glassy) energy landscapes.  A `benchmark' algorithm naturally suggested by
  this construction finds solutions in polynomial time up to a point beyond
  the `clustering' and in some cases even the `thermodynamic' transitions.
  This point has a simple geometric meaning and can be in principle determined
  with standard Statistical Mechanical methods, thus pushing the analytic
  bound up to which problems are guaranteed to be easy.  We illustrate this
  for the graph three and four-coloring problem.  For Packing problems the
  present discussion allows to better characterize the `J-point', proposed as
  a systematic definition of Random Close Packing~\cite{OHern}, and to place
  it in the context of other theories of glasses.  \vspace{1cm}
\end{abstract}


\maketitle

\section{Introduction}
\setcounter{equation}{0}
\renewcommand{\theequation}{\thesection.\arabic{equation}}
\label{Introduction}

Constraint Optimization and Satisfaction problems are a particular yet
widespread class.  The prototype is the Packing Problem, in which we are given
a fixed volume $V$ and asked to pack $N$ objects of typical size $r_o$ (a
scale factor) without overlap, making $N$ or $r_o$ as large as possible.  When
$r_o$ is increased (with $N,V$ fixed) the problem becomes harder as the number
of configurations that solve it decreases, until a point is reached
$r_o^{pack}$ beyond which there are no more solutions.

Another celebrated example is the satisfiability, or SAT,
problem~\cite{bookcomplexity}: we have a set of $N$ Boolean variables
$\{x_i=0,1\}_{i=1,\ldots,N}$ and a number $M= \alpha N$ of clauses ---when all
clauses have exactly $K$ literals, the problem is refereed to as K-SAT---
which in the present paper we shall assume are the first $M$ of a longer list
that is generated at random and stipulated once and for all.  We are asked to
find logical assignments satisfying
\begin{equation}
F=\bigwedge_{\ell =1}^M C_\ell =\bigwedge_{\ell = 1}^M\;\;\left(
 z_{i_1}^{(\ell )} \vee  
z_{i_2}^{(\ell )}\vee ... \vee z_{i_k}^{(\ell )} \right) \;\;\;,
\label{Fcnf}
\end{equation}
where $\bigwedge$ and $\bigvee$ stand for the logical AND and OR operations,
respectively, $\ell$ labels a set $i_i,...,i_K$
 and $ z_i^{(\ell )}$ is, depending on $\ell$ either $x_i$ or its
negation (this, and the $x_i$ that participate in each factor is what defines
the clause).  An assignment of the $\{x_i\}$'s satisfying all clauses is a
solution of the K-SAT problem.  When the number of clauses $\alpha N=M$
increased, the number of sets $\{x_i\}$'s satisfying all clauses decreases,
until a point $\alpha^{unsat}$ is reached beyond which there are no solutions.

The third problem we shall consider is the Graph $q$-Coloring
Problem~\cite{bookcomplexity}.  We are given a graph with $N$ vertices and are
asked to color them with one of $q$ colors so that no two vertices sharing a
link have the same color.  We shall assume as before that we have a list of
links, and we count the number of possible colorings when the graph has the
first $M=2 \alpha N$ of the list (where the average connectivity of the graph
is $c=2\alpha$).  As $M$ is made larger, a point $\alpha_{uncol}$ is reached in
which no more colorings are possible.

\begin{figure}[ht]
\begin{center}
  \includegraphics[width=.5\columnwidth]{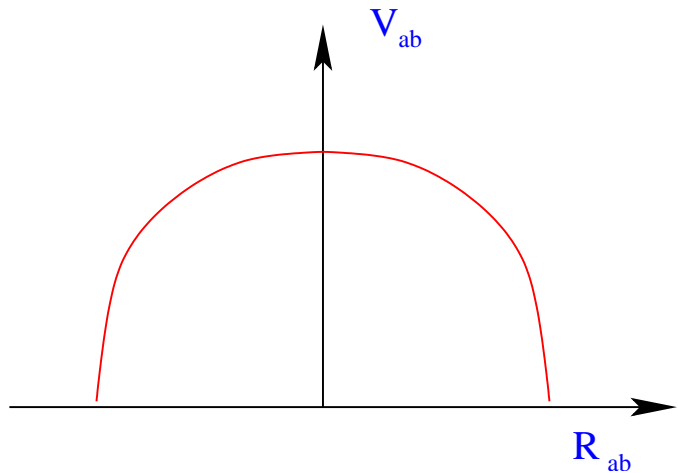}
\end{center}
\caption{Potential of soft particles with finite-range interactions. It allows
  to construct a packing version of the coloring problem (see text).}
  \label{fig2}
\end{figure}

Coloring problems are a particular class of packing problems.  Consider the
following `angle-packing' construction: in every node $i$ of a graph we assign
an angular variable $\theta_i$. There is a repulsive potential
$V(\theta_i-\theta_j)$ between the angles of linked sites as in Figure
\ref{fig2}, it is nonzero where the angle difference $(\theta_i-\theta_j)$
lies between $-2\pi/q$ and $+2\pi/q$ (modulo $2 \pi$).  For every angle
configuration with zero total energy, one can obtain a solution of the
$q$-Coloring problem just by assigning to the site $i$ the color numbered by
the integer $R= {\mbox {Int}} \left(\frac{q \theta_i}{2\pi}\right)$
($R=0,1,...,q-1$).  Conversely, each coloring solution has at least one zero
energy configuration counterpart.

Statistical Physics methods are particularly efficient in treating large
 random problems, especially  when mean-field techniques are exact, 
and indeed several important results where obtained for
 both satisfiability~\cite{SP,MZ} and coloring~\cite{Mulet} in this case.   
The connection with statistical mechanics is 
usually introduced as follows (see for example~\cite{Anderson,MePaVi,MZ}):
one first defines an energy $E$ such that it is zero when the problem is 
satisfied
and larger otherwise. In the examples above, this could be the number of 
overlapping
spheres, the number of unsatisfied clauses, and the number of links joining
 vertices
with the same color, respectively.
 Next, one studies for finite temperature $T=1/\beta$
the partition sum over configurations
\begin{equation}
Z=\sum_{conf} e^{-\beta E(conf)}
\end{equation}
The results  are encoded in the energy and entropy in the 
zero-temperature limit. One can apply all the analytical techniques
of Statistical Mechanics, and has in addition available all the standard 
practical methods
of annealing energies (in metallurgy as well as in computational physics) that
serve at least as a first, general purpose numerical tool. 

\begin{figure}[ht]
\begin{center}
  \includegraphics[width=.5\columnwidth]{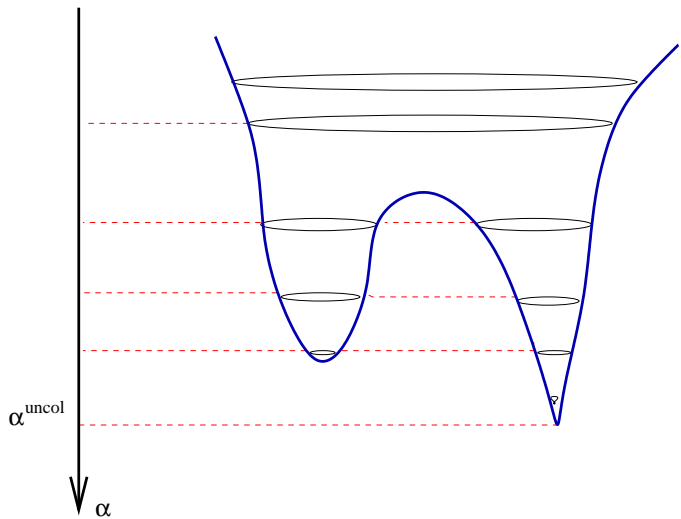}
\end{center}
\caption{Constructing a pseudo-energy from a constraint satisfaction problem:
the  regions shown are the satisfied sets at each level of difficulty.
The height function is such that at each level the intersection corresponds
to a constraint satisfaction problem.}
  \label{fig1}
\end{figure}

In this paper we shall follow a different strategy. We  first
introduce in section \ref{hypothesis} a 
pseudo-energy defined directly in terms of the Constraint Satisfaction problem.
For the packing problem, consider the set of configurations that satisfies the
 constraints at a given value of $r_o$. If we make the problem harder
by increasing  the common scale of all particles $r_o$, the set of
 configurations
still satisfying the constraints is strictly a subset of the previous one.
Thus, we can construct a single-valued 
landscape function ${\cal{E}}$ as in Fig \ref{fig1}. For the SAT and Coloring
problems we proceed similarly, with now $\alpha$  playing the role of
 $r_o$,
increased by adding clauses ({\it resp.} links) 
one by one following the predefined
 list. {\em The variable conjugate to  ${\cal{E}}$ is in fact a pressure}. 

Next, we  explore in section \ref{landscape} the
 consequences of making the following Landscape Hypothesis:
{\em The pseudo-energy  ${\cal{E}}$ landscape defined above has in the
 thermodynamic limit the same 
qualitative  properties  as the usual rugged energy
landscapes.} This means that pressure plays a similar role than that of an 
inverse temperature.
We shall find that several intriguing results \cite{all} in the geometry
of random Constraint Satisfaction Problems (CSP) such as the
satisfiability of random formulae or the coloring of random graphs
can be readily understood in this language. 

In section \ref{algo}, we describe a family of algorithms
 that immediately suggests itself: it 
is simply a quench in the pseudo-energy landscape, and hence by construction 
polynomial in the system size. At each step one increases the difficulty
of the problem, and moves to a configuration
in a finite neighborhood so as to restore satisfaction, if there is no
such configuration the program stops. We shall show that 
its performance --- where it converges and in what times ---
 can be in principle determined analytically for the very same problems for which
analytical statistical mechanical solutions are available.  
The method can be generalized to slower annealings, or to finite 
 pressure.

Next, we explore in detail the performance of this algorithm for the random
$q$-coloring problem. Although here we are more interested in constructing a
benchmark than a competitive numerical strategy, we find that the performance
is surprisingly quite good.

In section \ref{Jpoint} we concentrate on hard spheres and recognize that this
algorithm yields the procedure of O'Hern {\it et al.}~\cite{OHern} (and more
generally the Lubachevsky-Stillinger~\cite{LS} procedure if a slower annealing
is made) to define the so-called `J-point', which they propose to identify as
Random Close Packing.  Comparing with the random satisfaction problems of the
previous sections, we are thus able to give a mean-field realization of the 
J-point scenario\footnote{A true mean-field calculation for the J-point of
  hard spheres should also be possible\cite{Zampo}.}, and to put it in the context of other special points in glass theory,
such as the (putative) ideal thermodynamic glass state, and the
(finite-dimensional relic of) the Mode-Coupling transition. 

\section{Constraint Satisfaction Problems and 
a Landscape Hypothesis}
\label{hypothesis}

Let us describe the pseudo-energy in more detail. Consider a fixed set
of particles of any shape whose positions and angles
are given by the  coordinates ${\bf x}_1, ...,{\bf x}_N $,  
 with a  common scale $r_o$, such that for example
 multiplying $r_o$ by two doubles the linear size of all particles while
preserving their shape.
According to Fig. 1, the landscape at a point 
$ E({\bf x}_1, ...,{\bf x}_N) $ is defined as (minus) the first value  of $r_o$
at which there is a particle  overlap for these coordinates. (The minus sign is
 conventional).

In particular, for spheres of equal size $r_o$, the pseudo-energy
is simply (minus) the smallest inter-particle distance in that configuration
divided by two. As such, it is very close to the succession of potentials
 discussed by 
Stillinger and Weber~\cite{SW}:   
\begin{equation}
E_{SW}= - \sum_{a \neq b} \frac{1}{|{\bf x}_a-{\bf x}_b|^n} \;\;\; {\mbox{for}}
 \;\;\;
n \rightarrow \infty
\label{SW}
\end{equation}
Here we are considering
\begin{equation}
{\cal{E}}= - N \lim_{n \rightarrow \infty}
\left\{\sum_{a \neq b} \frac{1}{|{\bf x}_a-{\bf x}_b|^n}\right\}^{-\frac{1}{n}}
\label{uu} 
\end{equation}
(The factor $N$ is needed to assure that the quantity is extensive  if
we make the limit $n \to \infty$ {\em before} the thermodynamic limit.)
{\em Note that a gradient descent in (\ref{SW}) follows the same path 
as one in (\ref{uu})}. 

For the SAT and the Coloring problem the space is discrete, and the difficulty
of the problem may be increased by adding clauses and links, respectively.
The pseudo-energies are then, by analogy, minus the number of clauses, and
minus the numbers of links, respectively.  As mentioned above, we are
considering in fact a succession of problems, obtained by choosing the first
$M$ of a list of clauses or links which is fixed once and for all. Note that
if we could reshuffle the list, the net result at finite $M$ would depend on
the length of the list itself: for an infinite list we would optimize the
choice of $M$ constraints to make them easily satisfiable --- this is called
the `annealed' problem in the disordered system literature (not to be confused
with the `annealed approximation' of the probabilistic literature, which is
something else altogether~\cite{annealed} ).
 
 The landscape for the Coloring problem is very rugged indeed. 
 We have a graph with $M$ links that is well colored.
 Let us consider
 how the pseudo-energy changes when we flip the color of one given vertex.
 The new pseudo-energy is (minus) the largest value of $M'$
 that will make the new graph colorable, so we have to take away one by one
 the links {\em in reverse the order, starting from the last introduced} from
 $M$ to $M'$, but most of the links deleted will not be even close to the
 configuration we have flipped, and it will typically take $M-M' \sim O(N)$.
 We shall see that this poses no problem if one proposes changes only in
 nearby sites along the tree: thus, jumps will be always to
 configurations that are close and have lower pseudoenergy, avoiding other
 ones in the neighborhood that are much higher.

\subsection*{From constraint satisfaction to energy landscapes and back}

As mentioned above, in this paper we shall make the Landscape
Hypothesis that the pseudo-energy defined above has in the
thermodynamic limit the same qualitative features as expected from a
generic rugged energy landscape. We do not know if this is strictly
so, but it is clear that the converse is true: {\em every rugged
  energy landscape yields, in the microcanonical ensemble, a
  constraint satisfaction problem}~\cite{Moukarzel1}.  This is easy to
see just reading Figure \ref{fig1} in the reverse way: we imagine that
the landscape $H({\bf x})$ is given, and by cutting (microcanonical)
slices of fixed energy at different depths ${\cal{E}}$, we obtain the
CSP of finding ${\bf x}$ such that $H({\bf x})<{\cal{E}}$.

What we shall do in the following section is to briefly review some
known  facts from 
complex energy landscapes, and then proceed by cutting microcanonical
slices to infer the behavior it implies for random 
Constraint Satisfaction Problems.

\section{Energy Landscape: microcanonical slices}
\label{landscape}

Let us review what we have learned, mostly from mean-field
analytic calculations,  about rugged energy landscapes.
In order to match the constraint satisfaction situation, 
we shall adopt the rather unusual point of view of making energy slices,
rather than   constant temperature ones. The states we encounter shall then
be  regions of the energy shell that are dynamically isolated.
Figures \ref{sk}, \ref{pp} and \ref{pps} show various of these
constant energy slices: the temperature $T$  is then a variable that
classifies the states at that energy: in the landscape of
${\cal{E}}$ it amounts to classifying states according to their pressure
(the inverse of the temperature conjugate to ${\cal{E}}$).
 This is not the usual practice in optimization problems,
which is to classify the states according to their
internal entropy $\sigma$ (the logarithm of the number of configurations,
$\frac{1}{T}=\frac{\partial \sigma}{\partial E}$).

\subsection{Energy vs. Free energy barriers}

\begin{figure}[ht]
\begin{center}
  \includegraphics[width=.5\columnwidth]{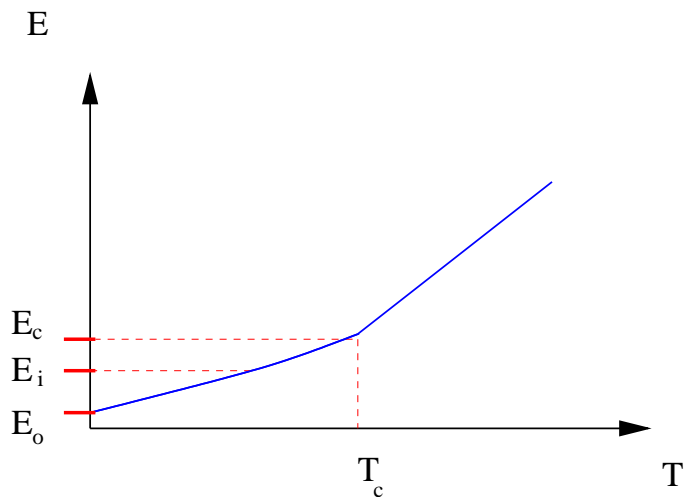}
\end{center}
\caption{Energy vs. Temperature of  a ferromagnet}
  \label{fig3}
\end{figure}

In what follows, we shall frequently allude to `states', defined as regions
of phase space that are dynamically isolated, in the sense that
  the  system starting a random walk within them will take a long time
to exit. `Long' might mean `infinite', or simply diverging with the system 
size $N$ faster than any power law.
Consider for example a ferromagnet with  continuous spins
on a lattice of dimension $d>1$ and linear size $L$, with
 nearest-neighbor interaction
\begin{equation}
E=- \sum_{\rm next~neigh.} s_i s_j - A \sum_i (s_i^2-1)^2
\end{equation}
In terms of the temperature, the phase diagram looks as Figure \ref{fig3}:
there is a critical temperature below which there are two states of 
magnetization $M = \sum s_i= \pm m L^d$, and above one state of zero
 magnetization.  
As is well known, a system at $T_i<T_c$ performing  Monte Carlo
dynamics
starting with magnetization density $m$ 
 will take a time 
to jump to the state of magnetization $-m$ that diverges in the thermodynamic
limit. 
We can also make a random walk {\em at constant energy} $E_i$, and
the time to pass from the $+m$ to the $-m$ state is still divergent.
 Does this mean that the constant energy surface $E({\bf s})=E_i$ is
 disconnected in two regions? A moment's thought shows that this cannot be 
the case. The actual saddle point in energy separating  the two 
energy minima 
is just the energy to create an interface
between half lattice at  magnetization $+m$ and half at $-m$,  
$E_{saddle}-E_o \sim  L^{d-1}$, while    $E_i-E_o=O(L^d)>>E_{saddle}-E_o$
 (see Fig. \ref{fig3}).
We conclude that the energy surface is actually guitar-shaped, and the reason 
why the system takes a divergent time to pass from one state to the other is
that (for large size $L$)  the neck is exponentially thin.

\subsection{Flat bottom states}

Consider a system of particles with a finite-range potential as in Figure
\ref{fig2}, confined to a fixed volume. To the extent that
the packing without overlaps is possible, it is clear that the total energy 
landscape will have minima of zero energy which are {\em flat}.
Surprisingly (and confusingly) also the pseudo-energy landscape can have 
`flat bottom' states, both in the particle and in the K-SAT and Coloring
problems, as we shall see in the next section.
 They turn out to be very important.

\subsection{Mean-field glass landscapes}

We shall now give three examples of mean-field landscapes with many states.
Mean-field  problems in physics correspond to large random problems
 in optimization and satisfaction: 
the coloring of large random graphs or the satisfiability of random
 formulas are instances of this class.
 Indeed these are the models for which we expect a correspondence between
 landscapes  to hold. 

The first example (the Sherrington-Kirkpatrick model)
 has the property that its  states having
free-energy density higher than the ground state one
are no obstacle for the dynamics. This class also includes all models
with smooth short range interactions in finite dimensions, like the
Edwards Anderson model.

The second example, the $p$ spin model is the precise opposite: a naive 
dynamics is trapped in a `threshold' level where the first, highest minima are
encountered --- this happens at energy density {\em above} the lowest.

The third example, the  Ising $p$-spin model, is the most typical of 
random optimization problems: it has a combination of `transparent' high states
that are avoided by the dynamics, but it also has a threshold level 
at finite energy density above the lowest, below which the dynamics 
only goes in times that diverge with the size.  

\subsubsection{The Sherrington-Kirkpatrick class}
The Sherrington-Kirkpatrick model~\cite{MePaVi} (SK)
 is the most studied spin-glass,
it consists of $\pm 1$ spins coupled by random, fully connected interactions
\begin{equation}
E=\sum_{ij} J_{ij} s_i s_j
\end{equation}

\begin{figure}[ht]
\begin{center}
  \includegraphics[width=.5\columnwidth]{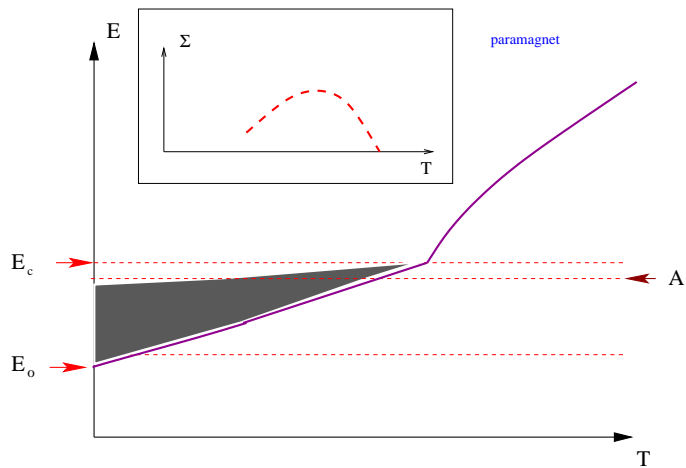}
\end{center}
\caption{A sketch of the metastable-state space of the Sherrington-Kirkpatrick
 model: the gray area in the graph corresponds to a region with an
 exponential number of metastable states. In the inset: complexity
 along a constant-energy surface line A in terms of temperature $T$
(the same graph would be obtained by plotting in terms of state entropy 
$\sigma$.)}
  \label{sk}
\end{figure}

It was soon realized~\cite{De,BM} that the free energy landscape had many
metastable states (see \cite{MePaVi}), a situation depicted in Fig. \ref{sk}.
The dynamics in the thermodynamic limit has been also solved~\cite{Cuku2}, and
surprisingly one finds that at each constant temperature the energy {\em
  density} tends for large times to the equilibrium one. In other words: {\em
  all but the very lowest metastable states are transparent to the dynamics},
if one wishes to compute the energy {\em per spin} with any percentage
accuracy, this can be done in polynomial time (although the true ground state
might might be harder to find).  It took over a decade to clarify completely
this from a pure `landscape' point of view~\cite{Blythe}.

 Because the states are not strong dynamic traps,
a  perturbation with small random forces that do not derive from a potential --
 a weak  `stirring' --- immediately sets the system into motion: the whole state structure
is washed away. 

\subsubsection{Spherical and Ising p-spin glass.}

\begin{figure}[ht]
\begin{center}
  \includegraphics[width=.5\columnwidth]{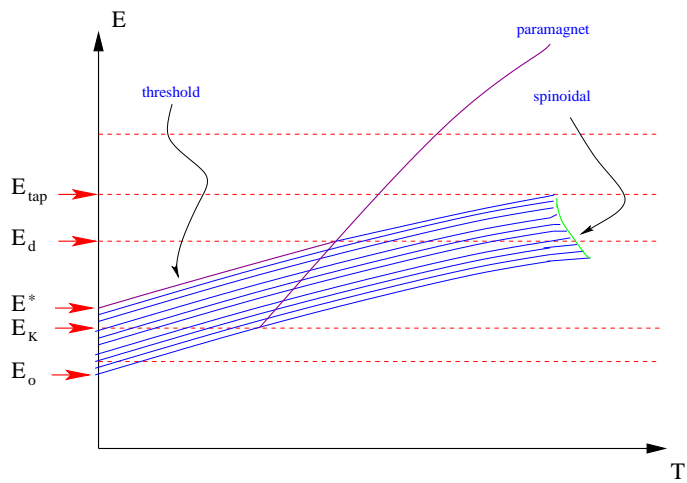}
\end{center}
\caption{The structure of metastable states of the spherical $p$-spin model.}
  \label{pp}
\end{figure}

The spherical $p$-spin model ($p>2$) has a Hamiltonian:
\begin{equation}
E=\sum_{i_1,...,i_p} J_{i_1,...,i_p} s_{i_1} ... s_{i_p}
\end{equation}
with the $s_i$ satisfying a spherical constraint.
It has a well studied 
\cite{KuPaVi,CrSo} structure of states shown in Fig. \ref{pp}.
Each line in the figure symbolizes a state (a free energy minimum). These
do not merge and keep their order. There is a {\em threshold} level 
above which all the stationary points are unstable.
States under the threshold are stable, the more so the deeper, and those
that are just below  are {\em marginal}, the 
`spin glass susceptibility' diverges within them~\cite{KuLa}.
 A system sitting on the threshold states will be set into motion by
small non-conservative `stirring' forces (another manifestation of their marginality),
but this is not the case with deeper states which are `rigid'.

The zero-temperature intercept of the threshold energy $E^*$
will be of particular
importance for us. Just above this point are the overwhelming
majority of energy barriers~\cite{KuLa}, and 
 this means, according to the discussion above,
  that for all energy slices 
 higher than $E^*$ the states  are in fact connected by narrow bridges. 

The out of equilibrium dynamics of this model is well studied~\cite{Cuku1}, 
 a quench starting from a random configuration never goes below
the threshold level. At zero temperature, a quench ends in the energy $E^*$.
 In 
 fact, no (instance independent)
external field protocol  is known that leads below the threshold:
temperature cycling, magnetic field changes,  
even quantum annealing~\cite{Let} have been tried, 
but the system always ends up on
the threshold level at the end of the cycle.

The thick line of Fig. \ref{pp} labeled `paramagnet' corresponds to
the equilibrium high-temperature phase. Above $E_d$ this is
a big state that dominates the dynamics and contains the overwhelming
majority of configurations. Between $E_K$ and $E_d$ the `paramagnet' is 
fractured in many states sharing each a negligible volume, the probability of
two configurations chosen at random of being in the same state is zero.
Below $E_K$  the lowest states dominate, there are still many states 
having a non-negligible volume, but the sharing of volume is so inegalitarian that
two configurations chosen at random have a finite probability of being
in the same state.  

\begin{figure}[ht]
\begin{center}
  \includegraphics[width=.5\columnwidth]{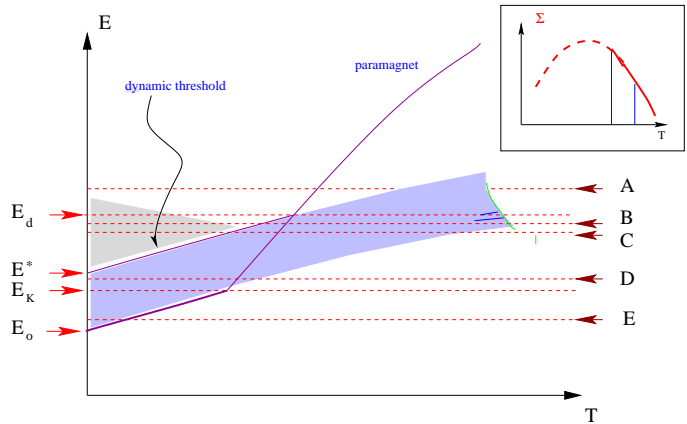}
\end{center}
\caption{A sketch of the organization of states in the Ising $p$-spin model.
In the inset: complexity in  slice C; the two vertical lines indicate the
 dynamic threshold (left) and the typical configurations (right), the
dashed line symbolizes states that are transparent to the dynamics.
The situation in slices A-E have been all found in constraint satisfaction
problems for different levels of constraints (see text).
}
  \label{pps}
\end{figure}

Microcanonical slices at different energy levels give us a first glimpse of
what happens in a constraint satisfaction problem. The number of states,
as calculated in \cite{CrSo}, grows with decreasing temperature (at constant
energy), up to $T=0$ ---
or  the intersection with the threshold line, if this happens first.
A system prepared at any energy $E^*<E<E_d$ performing energy-conserving 
dynamics~\cite{foot0} will end
in the threshold level. If, on the contrary, $E>E_d$ the dynamics will 
evolve in the paramagnetic state.   

Note the presence at $E_d<E<E_{TAP}$, in a range in which the paramagnet
dominates,  of states that are completely irrelevant
from the dynamic (or static) point of view: these where found in
\cite{all,all-col,all-sat} for the coloring and SAT problems in a given range of parameters.

The spherical $p-spin$ model is somewhat special, and in fact the generic 
situation is richer. For example, the Ising version of the $p$-spin
model has a structure sketched in  figure \ref{pps}.
In addition to the states that resemble those of the spherical case,
there is a set of high-lying states that resemble those of the SK 
model~\cite{AF} and
are transparent to the dynamics. A small `stirring' perturbation wipes out
the high-lying states, while the deep non-marginal ones are stable.
  There is still a dynamically defined
threshold level above which states are marginal~\cite{foot4},
 but it
 is still not clear if there is a purely `local' characterization of
it
(but at any rate an exact dynamic calculation is always 
possible~\cite{Cuku2}). We shall in the last section also propose a static
 -- though non-local -- calculation to achieve the same goal.

\begin{figure}[ht]
\begin{center}
  \includegraphics[angle=270,width=.6\columnwidth]{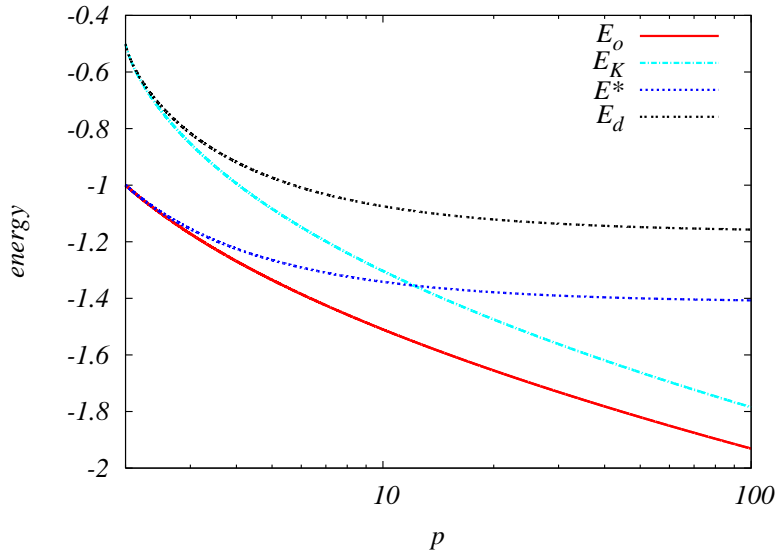}
\end{center}
\caption{The energies $E_d$, $E_K$, $E^*$ and $E_{o}$ for the $p$-spin model
  as a function of $p$ (from \cite{ALAIN}). A simple gradient descent program
  goes to $E^*$ which is beyond $E_d$, $E_K$ for $p<13$.  }
  \label{points}
\end{figure}

Making microcanonical sections as in figure \ref{pps}, we see some of the
possibilities for the organization of states and their respective dynamical
and equilibrium properties.  The inset of Figure \ref{pps} shows a sketch of
section along the line C: the states with smallest internal entropy correspond
to the `gray' region of dynamically transparent `SK-like' states.  This
feature has been found in K-SAT~\cite{Full_sat} and in the Coloring
problem~\cite{Full_col}.  The typical configurations (vertical line to the
right in the inset of Fig.\ref{pps}) are those that maximize $\sigma+\Sigma$
and do not coincide in general with the threshold states~\cite{MMM}.

In any case, the level achieved with a zero-temperature annealing $E^*$
may (and as we shall see will) be lower than $E_d$ and even $E_K$.
For example, in the $p$-spin model these points can be calculated analytically
\cite{KuPaVi,Cuku1} and the results are plotted in figure \ref{points}.

The important conclusion of this figure is that {\em if the task is
to find a configuration with energy as small as possible, generically 
neither $E_d$ nor $E_K$ pose a serious limit in themselves }, because $E^*$ may be smaller.
 {\em The same is true, for the same reason,  for constraint satisfaction problems}. 
This is the more so the smaller the values of $p$ -- and eventually
$K$ and $q$ in SAT and Coloring.
We shall see this  in the Coloring Problem in the next section.

\subsubsection{Random constraint satisfaction problems}
\begin{figure}[ht]
\begin{center}
  \includegraphics[width=1.0\columnwidth]{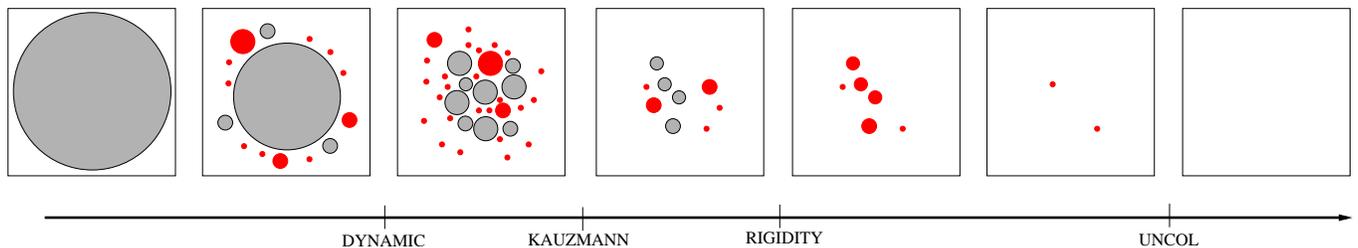}
\end{center}
\caption{Sketch of the set of solutions in random CSP when the
  connectivity is increased (adapted from \cite{all,all-col}).  From
  left to right, in order of increasing difficulty 1) easy problem,
  the whole solution space is connected, 2) a negligible fraction of
  solutions become disconnected, 3) `Dynamic transition' $\alpha_d$:
  the whole space breaks into dynamically isolated regions, two
  configurations at random belong to different regions, 4) Replica
  Symmetry Breaking, Kauzmann point $\alpha_K$: the regions become so
  rare that the probability of two configurations being in the same
  region is now non-zero, 5) Equilibrium `rigidity point'
  $\alpha^{hard}$. Typical configurations have frozen variables. 6)
  Best packing, unSat, unCol
  ($\alpha_{pack},\alpha_{unsol},\alpha_{uncol}$): the last
  configuration satisfying the constraints disappears. Red and grey
  regions symbolizes clusters with and without frozen variables,
  respectively (color online).}
  \label{CSP}
\end{figure}

The random satisfiability and the random graph coloring problem have been
studied using the cavity method \cite{SP,Mulet} and their landscapes when
varying the connectivity was characterized in \cite{all,all-col,all-sat}.  It
is interesting to observe that they follow our landscape hypothesis: they
behave indeed like microcanonical slices of the typical energy landscapes we
have been discussing. This is best seen in figure \ref{CSP}.

For the coloring problem, the col-uncol transition appears at
$\alpha_{uncol}(3)=2.355$ and $\alpha_{uncol}(4)=4.45$ ~\cite{Mulet}. The
other critical connectivities in the properties of the landscape were
found and computed recently by \cite{all-col}. Using 3 colors, one has
$\alpha_d(3)=2$ (a result already obtained in \cite{SAAD}), $\alpha_K(3)=2$ and
$\alpha^{hard}(3)=2.33$.  For the 4-coloring problem, the critical values are
$\alpha_d(4)=4.175$, $\alpha_K(4)=4.23$ and $\alpha^{hard}(4)=4.42$.

\section{Algorithms: arkless strategy for flood victims}
\label{algo}

Consider a rugged landscape that is being slowly flooded. We adopt the following
strategy: we stay still until the shore reaches our feet. We then move 
in small steps  
 away from the water, just enough each time to stay dry.
At a certain point, the patch where we are standing becomes an island, all land bridges
have been flooded: this is the `dynamic transition level'.
 We keep on moving uphill, just avoiding the waterline.
Our island may further divide into smaller ones, but our choice 
of island is dictated by 
the evolution of the shoreline.
 Our fate is sealed when our  island disappears altogether: this may happen
all of a sudden if its top is flat, or gradually if the top is rounded.
Note that nothing guarantees that we have ended in the highest 
summit, so our survival level is just locally --- but not globally --- optimal.
Clearly,  the  `dry land' is the  set of configurations satisfying the
constraints, and the `level of the water' the number of clauses, links or
particle size, which we increase gradually, while at the same time moving
in small steps to remain satisfied. The point reached with this algorithm is
image (5) of Fig. \ref{CSP}.

This is the `fast' procedure, a slightly more sophisticated one
is at each level of the water, not just to stay at the shoreline, but to explore
randomly all the available land, always without crossing water.
This means that on occasion we shall take advantage of a land bridge to go
 (randomly) from a promontory to another. 

A composite strategy is to explore all the available land at one level
of water, and then for all subsequent levels 
just follow the  direct 'lazy' strategy of moving just what is 
necessary.

Now, from the point of view of pseudo-energy landscape (which is reversed
with respect to the flood analogy, with summits becoming valleys),
 in continuous cases
it is clear that the `fast' algorithm is {\em just a gradient descent}, and
in general a zero temperature quench, while the  procedure in which we
take time to diffuse at constant energy, is a `slow annealing'~\cite{foot5}
in this case we have to specify how slow. The last, composite strategy,
is again a rapid quench, but this time starting from an equilibrium 
configuration
at some level.

\subsection{Hard sphere procedures.}

For the packing of hard spheres, the two procedures mentioned above have been
used by Lubachevsky and Stillinger~\cite{LS} (the slow annealing), and by
O'Hern {\it et al.}~\cite{OHern} (the rapid quenched). In both cases one
`inflates' all spheres simultaneously by a very small amount, and then uses
some repulsion to eliminate any overlap (wet toes, in the analogy above) that
might have been generated~\cite{Mak}.  In the annealed version, one also takes time to
diffuse without changing the radii or allowing overlaps.

Another possibility is to adapt the composite procedure used by Sastry et
al.~\cite{Sri} to analyze (true) energy landscapes: one starts from a fully
equilibrated configuration at a given (low) level of packing, and from there
onwards performs a fast quench, without diffusing at each step more than is
necessary to eliminate overlaps.

The reason why O'Hern {\it et al.} used the `rapid quench' rather than an
annealing is that their purpose was to define a point in parameter space (the
`J-point') in an unambiguous way. Had they used an annealing procedure, a
different point would have been obtained for each annealing time --- not to
speak about problems of crystallization in a mono-disperse sphere case (and,
indeed, the infinitely slow annealing limit would lead to a packing fraction
essentially corresponding to equilibrium).
 
As mentioned above, one of the aims of this paper is to put this `J-point' 
in the context of the rest of glass theory. We shall discuss this in 
further detail in section \ref{Jpoint}.
Let us note here that we can apply exactly the procedure of  O'Hern {\it et al.}
 (or  Lubachevsky-Stillinger) for the `angle-packing'
problem described in the introduction for a given, fixed graph,
 starting from $q=\infty$ and decreasing $q$. The smallest integer value of $q$ attained
gives then a realization of Coloring Problem. We shall not pursue this line here,
but rather attack the coloring problem from a different angle.

\subsection{Introducing temperature in pseudoenergy landscape. }

The pseudoenergy landscape suggests how to construct a Monte Carlo/Parallel
Tempering program for hard spheres.

Consider monodisperse hard spheres -- the generalization to polydisperse 
 is straightforward -- at `inverse temperature' (i.e.  pressure) 
 $\beta$.
Given a configuration, its pseudoenergy is proportional to (minus) 
the smallest 
interparticle distance $2 r_o$. We choose a particle at random and displace it
by a random amount. The minimal interparticle distance
 $2 r_o^{old}$ may have changed: if it is 
so it must be due to the distance between the particle just moved and 
one of its neighbors  becoming some value $r_o^{new}<r_o^{old}$
We thus accept the motion if $r_o^{new} \geq r_o^{old}$, otherwise we accept it
with probability $e^{-2\beta(r_o^{old}-r_o^{new}}$).
This is the Monte-Carlo procedure. One can now run several copies in 
parallel at different $\beta$ and implement the usual 
Parallel Tempering procedure.

 \begin{figure}[ht]
\begin{center}
  \includegraphics[width=.5\columnwidth]{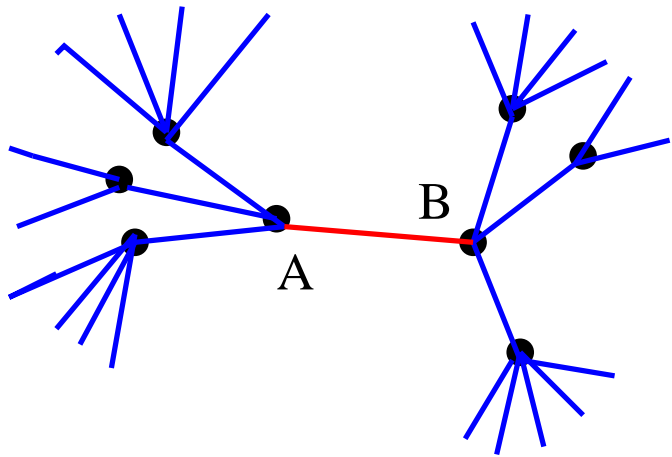}
\end{center}
\caption{Adding a link to a tree.}
  \label{tree}
\end{figure}
\subsection{Discrete lattice procedures}

We have chosen to exemplify the algorithm in the random $q$-coloring problem.
We are given a long list of links, and add them one by one.
 Suppose that 
we have a  graph
with $M-1$ links, and a configuration of colors for the vertices
satisfying the constraints. We now add the $M-th$ link from the list:
with probability one it will locally look as in Fig. \ref{tree}.
As we add the link, it may be that vertices A and B are of different colors,
in which case we proceed. If, on the contrary, the colors of A and B coincide,
we have to modify at least one of the two. 
This will typically create problems in the tree that has roots in them, so we
shall need to modify colors up the branches. If this can be done with a finite
number of trials, we do it, compute the total number of steps it took, and
proceed with the next link addition.  In practice, for the readjustment of the
trees with root A or B, we use a {\em Walk-COL} algorithm introduced in
\cite{all-col}, which is the coloring version of the celebrated {\em Walk-SAT}
program~\cite{walksat,foot2} This is perhaps not ideal, but it definitely
provides an upper bound of the number of flips needed. Moreover, this strategy
is straightforward to implement numerically. An important property of this
algorithm is that it is {\it focused}: it changes only frustrated variables
and it thus acts locally starting from the initial frustrated link, and does 
not perturb the solution far from this link.

In this form, the algorithm is a {\em recursive version}
 of the method called {\em Incremental-SAT} \cite{i-sat}
 in the
 computer science community, in which
one starts from a SAT formula and its known solution 
(or a graph and its proper $q$-coloring) and then 
adds one formula (or a new link for coloring) and has to  find a
 solution to the new problem. This is usually done using Walk-SAT, as we do, although
 this procedure has not been proposed as far as we know 
as a general way to obtain solutions starting
 from scratch. The point reached by our algorithm can thus be seen as a EASY/HARD 
transition for the recursive, incremental COL problem.

\subsection{Coloring random graphs}
\begin{figure}[ht]
\begin{center}
  \includegraphics[width=.5\columnwidth]{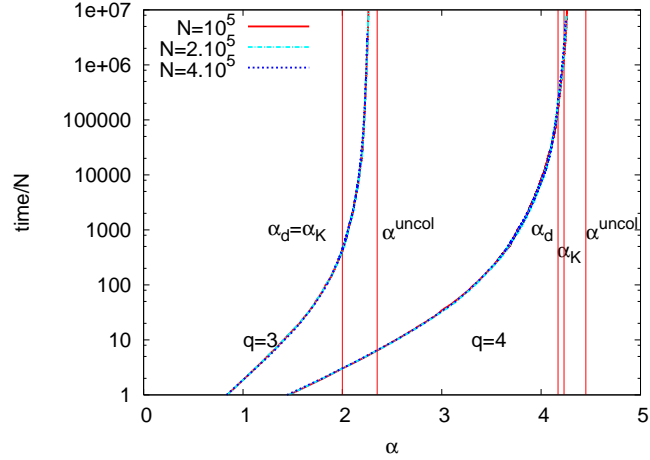}
\end{center}
\caption{Performance of the recursive algorithm for the $q=3,4$-random
  coloring problem. The time needed to find a proper coloring diverge at a
  connectivity $\alpha_*<\alpha_{uncol}$ in both case, but the algorithm is
  able to find colorings in linear time beyond the clustering $\alpha_d$ and
  Kauzmann $\alpha_K$ transitions.}
\label{diverge}
\end{figure}

We now discuss our results and show our data for the $3$-coloring and the
$4$-coloring in fig \ref{diverge}.  It turns out that the number of steps
needed (the number of re-colorings close to A or B to get back to satisfaction)
grows on average and diverges at a given connectivity $\alpha^*$, where our
program stops.

The fact that the rearrangement needed grows with the connectivity is not
surprising in view of the results of Montanari and
Semerjian~\cite{MS,recolorings}, who showed that the minimal number of
rearrangements needed to satisfy constraints following a change of color of a
random vertex diverges like a power law at a certain connectivity. The
distance along the graph of the rearrangements is a power law as well.  
This critical connectivity corresponds in fact 
to the rigidity transition $\alpha^{hard}$
 considered in \cite{all-col} where frozen variables (or `hard fields' 
in the cavity
language) appear that fix the value of a finite fraction of
variables for all configurations within a state.
  After a situation with frozen variables is reached, an additional link
will have a finite probability of connecting 
two  variables frozen to the same color,
thus rendering unsatisfied all configurations within the state.
States with hard fields have a high mortality rate:
after an extensive number of link additions, the survival probability
of such a state  is exponentially small in $N$. 

In fact, here the situation is somewhat different to the one 
in ~\cite{MS,recolorings}, in that we are not
considering an equilibrium configuration, but one obtained by a succession of
link additions and rearrangements.
 The actual value $\alpha^*$ that our program can
reach need not coincide exactly with $\alpha^{hard}$, because hard fields appear
at somewhat lower connectivities in our out of equilibrium procedure.

It can be shown
that for large connectivity, all clusters have frozen variables 
beyond a certain
connectivity (see for instance \cite{all-col,all-sat} or \cite{FEDE} for
rigorous proof in SAT): this puts a strict limit to the algorithm. For the $3$
and $4$ coloring problem, it has been shown (see again \cite{all-col}) that
all states have frozen variables for $\alpha \geq \alpha^{hard}$.

\begin{figure}[ht]
\begin{center}
  \includegraphics[width=.49\columnwidth]{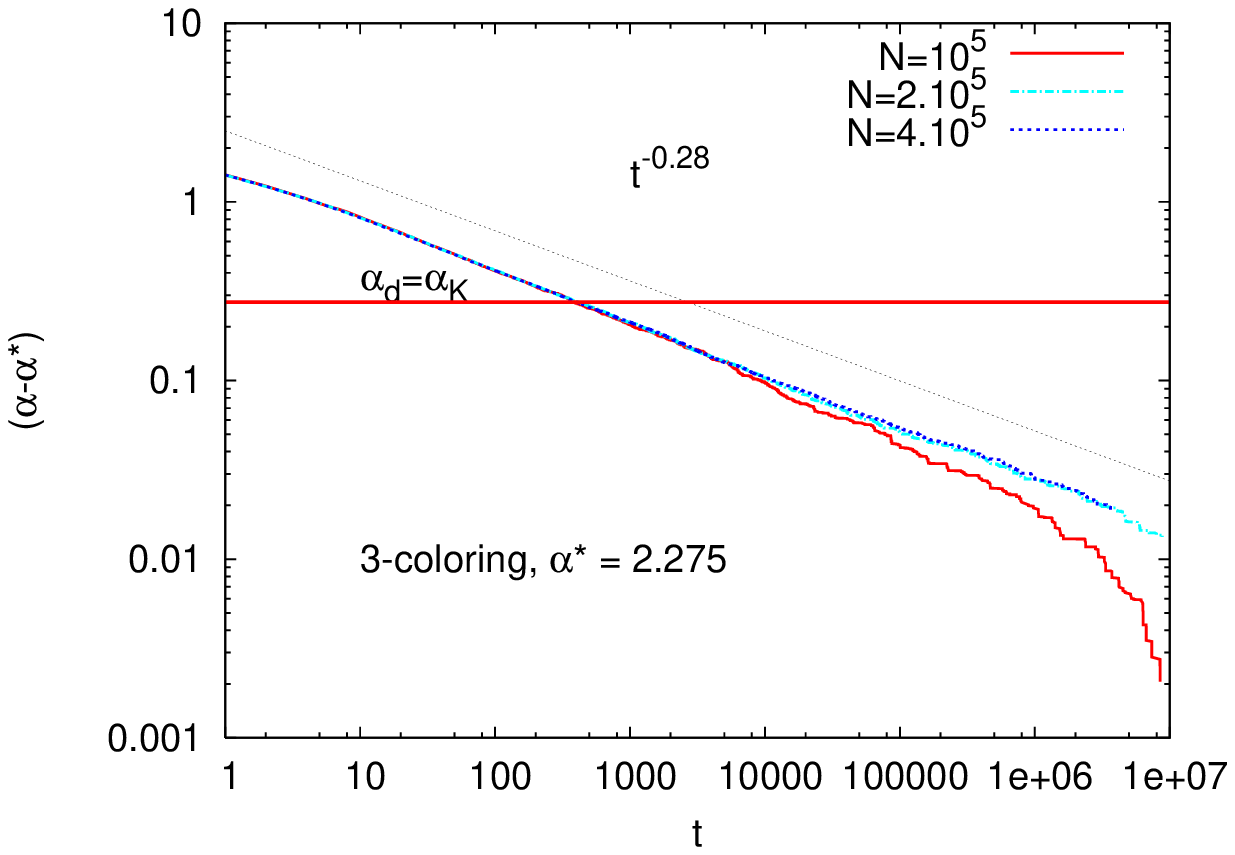}
  \includegraphics[width=.49\columnwidth]{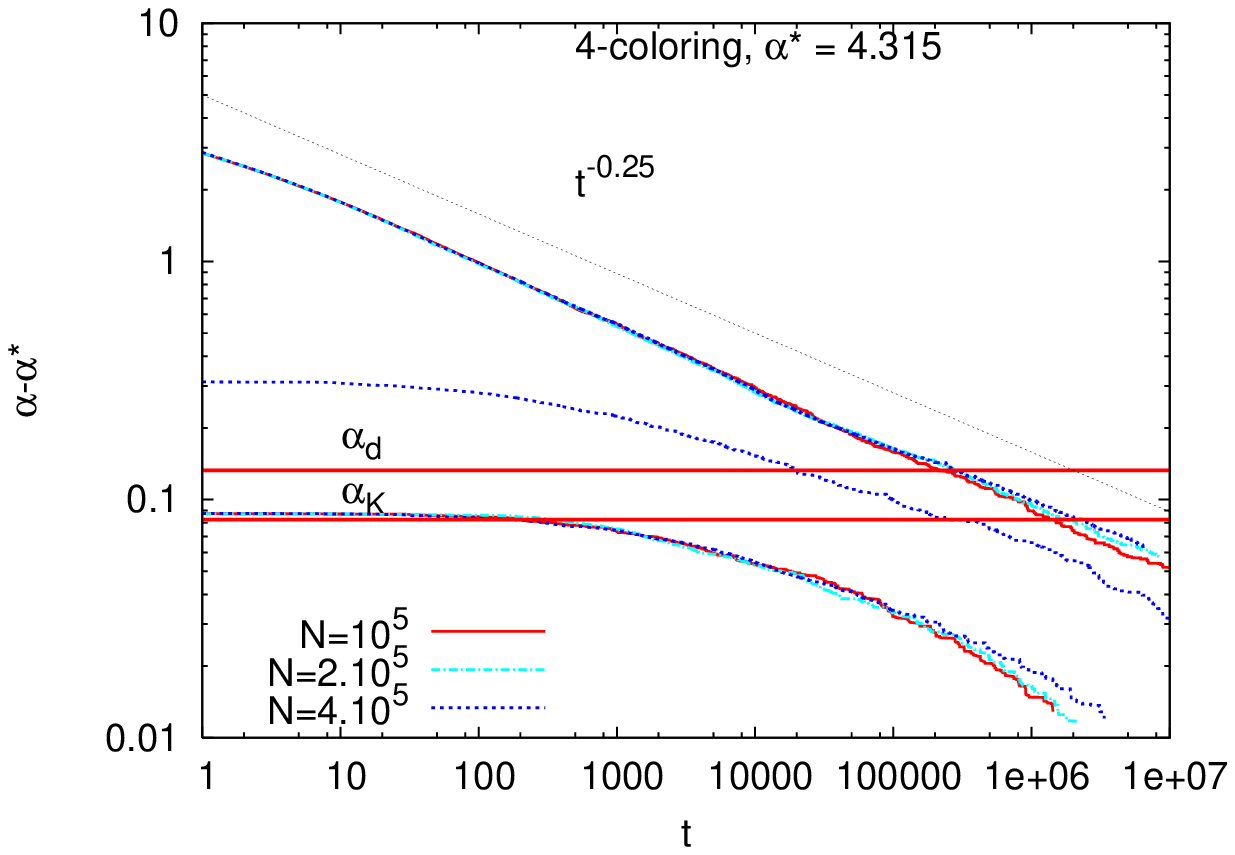}
\end{center}
\caption{Estimate for the asymptotic $\alpha^*$ of the algorithm. 
  Left: $q=3$-random coloring problem.  Right: $q=4$-random coloring problem.
  The two lower families of curves correspond to procedures starting from
  configurations obtained using the Belief Propagation algorithm of
  ref\cite{all-col}: the initial advantage seems eventually lost.}
  \label{q3q4}
\end{figure}

Because we can scale the curves with system size, and we consider only
the regions where the curves superpose properly, we can in fact infer
the behavior for $N = \infty$. We do that in fig. \ref{q3q4}.  In the
$3-$ coloring, the algorithm reaches in time linear in $N$ a value
$\alpha^* \approx 2.275$ which is beyond the clustering and Kauzmann
transition $\alpha_d=\alpha_K=2$. It is however, as expected, below
the appearance of frozen variables in the thermodynamic states that
take place at the rigidity transition $\alpha^{hard}=2.33$ (and of
course below $\alpha_{uncol}=2.345$).

In the $4$-coloring, the algorithm reaches in time linear in $N$ a value
$\alpha^* \approx 4.31$ which is again well above $\alpha_d=4.175$ and
$\alpha_K=4.23$ (but still systematically below $\alpha^{hard}=4.42$ and
$\alpha_{uncol}=4.45$).

The integrated number of steps up to connectivity $\alpha$ is
$(\alpha^*-\alpha)^{-\mu}$ with $\mu \sim 0.28$.  In Figure \ref{q3q4} we show
for reference the case $q=3$, which is somewhat more dubious because it is not
clear yet to which class -- from the replica theory point of view -- it
belongs.  In any case, the value $\alpha^*\sim 2.275$, far beyond
$\alpha_d=\alpha_K=2$, is quite close (but still smaller) to the one
($\alpha_{SP}=2.3$) obtained in Ref \cite{Mulet} with the best algorithm
---Survey Propagation (SP) \cite{SP}--- for smaller sizes.

Before concluding this section, we wish to compare for this case the
present algorithm with the `Belief Propagation' (BP)
algorithm~\cite{all,all-col}.  We start with a configuration obtained
with BP at a given $\alpha$ (which we shall try to make as large as
possible~\cite{BPalone}, and then improve the procedure by adding links one by
one.  The result is shown on the right of \ref{q3q4}: at the beginning
there is a very fast progress, but the asymptotic $\alpha^*$ is not
significantly different -- at least within our precision. We conclude
that starting from a BP solution does not seem to improve the scaling
with the system size.

To explain this, we can think in the flood analogy of BP as
 starting us somewhere in the middle
of an island. As the level of water keeps on going up, at the beginning
we have little to do to stay dry, and the program proceeds rapidly.
Only when the many-dimensional shore reaches us do we begin to move in a 
complex manner, and the system slows down.

\subsubsection*{Scalings with system size}

By performing at each step Walk-COL starting from a site at the end
 of a new link,
we have used  on average a certain function  $G(\alpha^*-\alpha)$
color  flips each time we add one link,
a quantity of order one.
The program `time' $t(\alpha)$ 
(the total number of steps up to connectivity $\alpha$)
 thus increases by  $G(\alpha^*-\alpha)$ 
each time that $\alpha$ increases by $1/N$:
\begin{equation}
 \frac{dt}{d\alpha}= NG(\alpha^*-\alpha)
\end{equation}
and $t(\alpha)$ is $N$ times the integral of $G$. Thus, the curves
$t(\alpha)/N$ for different $N$ should collapse, which we check in Fig.
\ref{diverge}.  The fact that the total time scales with $N$ holds to the
extent that the number of rearrangements is, for large graphs, independent of
the graph size.
 
Note that nothing guarantees that $G(\alpha^*-\alpha)$, defined on the basis
of Walk-COL, is the minimum average number of flips needed to make the
configuration satisfied after adding a link, but just an upper bound. On the
other hand, in a locally tree-like structure, the maximal number of moves with
any program is to try all the flips of sites at distance along the graph
smaller or equal than $G(\alpha^*-\alpha)$, that is $\sim
\alpha^{qG(\alpha^*-\alpha)}$, still a quantity independent of $N$.  Thus, the
fact that the local structure of the graph is tree-like has helped us to
obtain a program linear in $N$.

If we did not know that we had a tree-like structure --- for example if some
loops are present --- we would have to explore {\em all} configurations in
phase-space with a given number of flips, trying all the phase space points at
progressively larger phase-space distances until satisfaction is restored.  In
this way, $G(\alpha^*-\alpha)$ flips would be obtained after at most $\sim
N^{qG(\alpha^*-\alpha)}$ trials, and the total integrated time would be:
\begin{equation}
t \sim A N \int_0^\alpha d\alpha'\; N^{qG(\alpha^*-\alpha')} \sim \left(\frac{AN}{\ln N}\right)
 \frac{ \; N^{ qG(\alpha^*-\alpha)}}
{ qG'(\alpha^*-\alpha)}
\end{equation}
This is the absolute worse the program will do for the random graphs discussed
here, not using either Walk-COL or the knowledge of the local tree-like
structure.  Below $\alpha^*$, the performance is still a power law in $N$.

\subsection{Marginality, hard fields and rattlers.}
\label{marginality}

As we have seen in the previous sections, when we add links to the Coloring
Problem (or clauses to the SAT problem) each time the number of changes needed
 to keep the system satisfied on average increases. This happens because
as we approach the point $\alpha^*$ larger and larger clusters of vertices
`have to be moved together', changes that keep the system satisfied have to be
done in a correlated manner in each cluster.  The same happens with the
sphere-packing problem: the clusters of particles that are in contact, and
have to be displaced together at each step, increases with packing and at the
critical value of $r_o$ the whole structure percolates.  By the time
$\alpha^*$ (or the J-point, in a packing problem) is reached, the correlation
diverges, and the algorithm virtually stops. Some important remarks are:

{\em i)} the density of links at which this happens {\em depends on the
  procedure}, the point $\alpha^{hard}$ obtained at equilibrium 
need not coincide
with the one obtained through the sequence of additions of links we have been
doing here.  (This is why in Figure \ref{q3q4} we have tried to compare the
divergence obtained by addition of links starting from low link density, with
the divergence obtained starting from a very connected graph colored with a BP
program.  Admittedly, the difference, if any, is not large).  For a particle
system, the packing fraction at which this percolation happens also {\em
  depends on the procedure}.

{\em ii)} for the overwhelming majority of
runs of the algorithm, 
when the level $\alpha^*$ is reached, the addition of an extensive
number of links requires a divergent number of re-colorings. 
However, it will
generally happen that a finite fraction of vertices can still be recolored in
many ways, without destroying the satisfaction. For example, the intermediate
vertices in a short loop made of a succession of vertices of connectivity two
can be recolored in many ways. Similarly, when a particle system reaches
$r_o$, there still are particles, or groups of particles (the `rattlers')
which can move inside cages made by the others.

To summarize: the SAT, Coloring and Packing problems all have critical
situations that are {\em process-dependent} in which a subset of the
elements form a `backbone' subjected to `hard fields' from their
neighbors and `rattlers' that have still freedom. {\em This point
  corresponds to the EASY/HARD transition for the Recursive
  Incremental SAT (or Col) algorithm described here}. If we look at
the pseudoenergy of a state, with large probability the number of
configurations at the critical level of difficulty is still large (due
to the freedom of rattlers), but it disappears completely at a level
of difficulty just higher.  In other words: seen from the point of
view of the pseudo-energy, the minima have a `flat bottom' consisting
of the rattler configurations
\footnote{The actual density of rattlers and its
dependence on the annealing procedure for hard spheres 
is still a matter of debate.}.

\section{The J-Point}
\label{Jpoint}

In a series of papers \cite{OHern,TheseM} the highest density configuration of
hard spheres reached by `inflating' as in the section IV A have been
studied~\cite{foot1}. The point where this happens has been christened
`J-point'. As we noted above, they did not consider a slow annealing, as in
the Lubachevsky-Stillinger procedure, because this would make the system
evolve in an annealing-time dependent manner, ultimately allowing it to
crystallize.  For spheres in three dimensions the J-point packing fraction
turns out to be very close to the one quoted as `Random Close Packing'~\cite{Mak}, and
whether one can identify J-point and Random Close Packing is a debated
question~\cite{debate}.  The interest of the J-point lies at any rate in the
fact that it is {\em critical}, it has associated with it a diverging length
and soft modes \cite{OHern,Wyart}.  There is also good evidence that for
spherical particles it is {\em isostatic}, i.e.  that the number of contacts
is just the number it takes to immobilize the system, without redundant static
conditions.

A question that immediately arises is whether the J-point packing corresponds
to the `ideal glass' phase, that is, the best packed amorphous state {\em
  'without crystallization'}.  Putting the question this way, we immediately
run into difficulties because by considering different degrees of
crystallinity, we can obtain a continuum of denser packings \cite{Sal1}. The
random models studied here allow us to avoid -- at least to fix ideas -- the
conceptual complication of crystallization, and to concentrate on the glassy
aspects~\cite{crysta}.  We shall hence first discuss the SAT and Coloring,
which by construction have no ordered states.  On the latter the procedure of
Ref. \cite{OHern} can be repeated in the `angle packing' version without
modifications.

In SAT and Coloring we have already pointed out that the recursive Incremental
procedure of adding difficulty in a constraint satisfaction problem yields a
critical point with a diverging length, given by the range of rearrangements
necessary each time to satisfy the constraints 
(see Ref. \cite{MS,recolorings}).  This
point, which we identify as the {\em mean-field version of the J-point}, also
corresponds to the appearance of frozen variables in the state~\cite{ccc}.

In the framework of the pseudo-energy landscape the procedure leading to the
J-point amounts to a deep (zero temperature) quench in this complex landscape,
starting from a random configuration.  In this language, it is an 
  infinite temperature inherent structure.  Clearly, there is no reason for
this procedure to converge to the (glassy) ground state (otherwise it would be
the ultimate solution to the coloring and SAT problem): minima with the
largest basin of attraction are generically not the deepest.

Running the program several times starting from different initial conditions
leads to different end densities \cite{OHern}, distributed around a typical
value, the closer the larger the system: The pseudo-energy view allows us to
recognize this as the standard situation with the inherent structure energy
distribution~\cite{walter}.

Another thing to notice is that configuration reached in the J-point is {\em
  not} typical of the given packing fraction, just as an inherent structure
reached by a quench in an energy landscape is not the typical one of the
corresponding energy.  One way to emphasize this is to envisage the analog
of the protocol of Ref.~\cite{Sri}: starting from {\em equilibrium}
configurations at different initial packing fractions (or values of $r_o$),
one applies the inflating procedure of O'Hern {\it et al.}  If the landscape
hypothesis advocated here holds, the final packing fraction (or $r_o$) reached
should depend on the initial one in much the same way as in Ref. \cite{Sri}
the inherent structure energies depend upon initial equilibrium energies (in
finite dimensions crystallization effects have to be taken care of
~\cite{crysta}).  It would be interesting to study other amorphous
configurations with the same packing fraction as the J-point, but reached
after different annealing protocols, and check whether they are not critical.
This is what happens in the mean-field models discussed above: their {\em
  typical} configurations at a packing fraction equal to that of the J-point
are {\em not} critical, even in this situation in which there is no
possibility of crystalline order.

\section{Analytical
 computations of  performance.
 Dynamics and Causal Replica Chains.}
\label{analytic}

The statistical mechanical description of complexity, in particular based on
the replica trick, concerns the average geometric structures in a given subset
of phase-space. This information is local, in the sense that it involves the
structure of typical states and not their whole basin of attraction.
Sometimes, this local information is enough to infer where a dynamic process
starting from a random configuration will go. This is the case for example of
dynamics of certain mean-field models, where `marginal' states are chosen.
However, it is not clear in general up to what point local information of
states is sufficient.  One alternative that has been widely discussed in the
literature, is to solve the averaged dynamics (for a review of this approach,
see: \cite{Leticia}). Here we shall discuss a different but related approach,
which is a generalization of several works in the replica literature of the
80's and 90's \cite{PaVi,Remi,FrPa} and in optimization problems \cite{MMM}.

Let us first consider an energy landscape of discrete variables.  We start by
choosing a random configuration ${\bf s_1}$ (here we use boldface to denote
$N$-dimensional vectors).  Next, keeping this one fixed, we search for another
configuration ${\bf s_2}$ at distance $\Delta_1$ from the first. Keeping these
two fixed, we choose a third configuration ${\bf s_3}$ at distance $\Delta_2$
from the second, and so on.  In this way, we construct a `causal chain', since
the subsequent links do not affect the preceeding ones.  We have to demand,
for example, that each new replica be at an energy as small as possible
(subject to its constraints): this is done by thermalizing each link at a very
low temperature.  Next, we have to make the distance between links tend to
zero, and the number of links to infinity.  From the point of view of replica
theory, this is just a generalization of the `effective potential' method
\cite{FrPa}, and is also closely related to the discussion in Refs.
\cite{PaVi,Remi}.  We have to write a partition function:
\begin{eqnarray}
Z &=&\lim_{n_1 \rightarrow 0} \ddots \lim_{n_R \rightarrow 0} \;
 \int d{\bf s_1^{\alpha_1}} d{\bf s_2^{\alpha_2}}
\ddots d{\bf s_R^{\alpha_R}}
\;\; {\bf \delta}({\mbox{chain}}) \nonumber \\
& & \exp \left\{ -\beta_1 \sum_{\alpha_1=1}^{n_1} E({\bf s_1^{\alpha_1}})
 -\beta_2 \sum_{\alpha_2=1}^{n_2} E({\bf s_2^{\alpha_2}}) -\ddots
 -\beta_R \sum_{\alpha_R=1}^{n_R} E({\bf s_R^{\alpha_R}}) \right\}
\end{eqnarray}
Where
 ${\bf s_1^{\alpha_1}}, {\bf s_2^{\alpha_2}} \ddots {\bf s_R^{\alpha_R}}$
are all $N$ dimensional vectors of spins.
 ${\bf s_1^{\alpha_1}}$ is replicated $n_1$ times,  ${\bf s_2^{\alpha_2}}$
 $n_2$ times, etc.
 Spins are organized in a hierarchical manner:
  one of the $n_1$ replicas  ${\bf s_1^{\alpha_1}}$ has an
offspring of
$n_2$  of the  ${\bf s_2^{\alpha_2}}$. One the $n_2$ of these has
in turn an offspring of $n_3$ of the ${\bf s_3^{\alpha_3}}$, and so on.
The term $ {\bf \delta}({\mbox{chain}})$ is a product of deltas imposing
that the $r$-th   `offspring' is at a fixed distance  $\Delta_r$ from its parent.

For a constraint satisfaction problem, the problem can be recast as follows.
 Taking the example of the  coloring problem, we  
suppose we have
a graph with  connectivity per node $\alpha_{max}$.
 The links are given by an $N \times N$
symmetric matrix $J_{ij}$ with elements one and zero. An ordering is introduced by
defining the symmetric matrix $C_{ij}$, with elements taken at random with
 distribution uniform
in $(0,\alpha_{max})$. The number of miscolorings at connectivity $\alpha$ is then:
\begin{equation}
 M({\bf s},\alpha)= \sum_{ij} J_{ij}  \Theta(\alpha - C_{ij})  \delta_{s_i s_j}
\end{equation}
where $\Theta(x)$ is one if $x$ is positive, and zero otherwise.
One repeats now the argument as before, for the case of a chain of replicas with 
progressively
larger values of $\alpha$, keeping the total number of errors equal to zero.

In practice, one can make the computation for   chains of a small numbers of links,
and extrapolate the result to the limit of continuous chain.
For systems on the Bethe lattice, the replica calculation should be translated
into the cavity language, which is more flexible. 

\section{Conclusions}
\label{conclusion}

We have adopted a unified point of view of hard-particle glasses and the
constraint-satisfaction problems of computer science and taken both to a
(pseudo) energy landscape setting.  On the one hand, the glass literature has
long been a source of inspiration and methods to attack satisfaction problems.
On the other hand, the statistical mechanics of optimization problems provides
nontrivial yet solvable models for which questions on the jamming transition
can be answered exactly.

In both cases, there are questions ---
such as a rigidity (hard field) threshold, the presence of `rattler' 
 and a backbone of fixed particles  (or Boolean variables, or colors)  ---
 peculiar to the fact that the task 
is posed as a zero temperature problem of optimizing  parameters 
while respecting 
a constraint. 
The pseudo-energy language helps to bring back many of these results
to  a problem of relaxation in a complex landscape.

\subsection*{Satisfaction}

The statistical mechanic point of view of satisfaction has been mainly
concentrated on computation of static `snapshot' properties of
phase-space. This approach has been very fruitful and has suggested
powerful algorithms \cite{SP}.  Here, instead, we take the approach of
defining an algorithm whose properties can in principle be computed
analytically, at least in the same cases for which a static solution
is possible.  This change of perspective reflects a similar passage
from a static to a dynamic approach in the glass
literature~\cite{Leticia}, and has also been pursued in the
statistical mechanics of optimization ~\cite{ReSi}.  In this paper we
have only stated but not completed the analytical calculation, but we
have shown numerically that problems are easy beyond the `clustering
transition', and with the Coloring Problem that in some cases even
beyond the `Kauzmann' one.  Indeed, one of the conclusions is that a
rather excessive importance has been attributed to the clustering
(dynamic) transition point: this is the point up to which any program
can easily sample {\em all} configurations~\cite{uff}, but not the
last point at which it can easily obtain at least {\em one}
configuration --- this happens at $\alpha^*$.  It is then, upon
reflection, not so surprising that relatively straightforward
algorithms such as Walk-SAT \cite{Asat} find easily solutions almost
up to the satisfiability threshold.

\subsection*{Jamming transition}

When the J-point was introduced~\cite{OHern} and shown to correspond
to a packing fraction close to the empirical value of `random close
packing', the question arose of its meaning in terms of the glass
state. What the present approach underlines (but was already implicit
in \cite{LS}), is that the nature of this point is the same as that of
a zero-temperature quench in a rugged energy landscape~\cite{foo}.
Deeper, more compact, points can be reached with other annealing
procedures - {\em even in models without a crystalline state }, as the
examples we have considered show.  This suggests a picture of the
transition in the temperature/shear-rate/density space \cite{LN},
rather than as a single surface, as a multi-layered onion, each level
being reached after a different protocol.

There is density at which a hard particle system has an experimentally
unaccessible relaxation timescale: this happens at a density
distinctly smaller than the J-point one. Indeed, the present analysis
shows that the J-Point is necessarily denser than the Mode-Coupling
transition density, in models in which such a transition exists and is
sharp. On the other hand, the question of which point is denser, the
J-Point or the Kauzmann transition point depends on the system and the
dimensionality -- again based on what can be inferred from models for
which the latter transition can be shown to exist \cite{Zampouno}.
 Note that, if the 
J-point is denser than the glass transition point, this is at the
expense of being out of equilibrium.
   
An interesting aspect of the J-point is that it is critical. Here we
have found that the same is also true for mean-field models (as had
already been found in the $q$-core problem \cite{perco}), but this
criticality does not correspond to the thermodynamic glass transition
itself (which in mean-field exists), but is that of the
zero-temperature `threshold' states~\cite{Cuku1} already familiar from
the mean-field glassy dynamics. In fact, in any of the problems
discussed here, if the procedure of approach to the J-point is stopped
when the correlation length has reached a certain large value, and a
thermalization at fixed $r_o$ (or $\alpha$) is applied subsequently,
as the system tends towards equilibrium the pressure drops and the
correlation length {\em diminishes}, rather than growing.

The study of problems with disorder raises doubts on 
the fact that  isostaticity of a frictionless particle system 
is a necessary (apart from sufficient ~\cite{Moukarzel,TheseM}, although see
~\cite{Zampo})
condition for  criticality. This may be checked just by 
testing the J-point procedure with a packing
of non-spherical (e.g. ellipsoidal) 
particles, which are generically hypostatic ~\cite{hypo}:
if the J-point for ellipsoids turns out to be  critical,
 then it will be clear that isostaticity
is not a necessary condition for criticality.

Finally, the analogy between particle and satisfaction problems suggests that
one may perhaps build a mean-field, `Random First Order' theory of rigidity
(and to a certain extent plasticity) by considering systems which develop
`hard fields', and subjecting them to driving, nonconservative forces.  The
Coloring Hamiltonian, or its `angle packing' version, is a good candidate for
such a strategy.

{\bf Acknowledgments} We wish to thank B. Chakraborty, S. Franz, A. Liu, M.
M\'ezard, R. Monasson, A.  Montanari, F. Ricci-Tersinghi, S. Sastry, G.
Semerjian, L.  Zdeborov\'a and R.  Zecchina for discussion about these issues.

\pagebreak


\begin{thebibliography}{99}


\bibitem{OHern}C. S. O'Hern, L. E. Silbert, A. J. Liu, S. R. Nagel,
 Phys. Rev. {\bf E 68}, 011306 (2003)

\bibitem{Mak} A compression procedure closely related to
the experimental one, and yielding densitites close to random close packing  is:
 Hern\'an A. Makse, David L. Johnson, and Lawrence M. Schwartz,
Phys. Rev. Lett. {\bf 84}  (2000)

\bibitem{bookcomplexity} M. Garey and D. S. Johnson,
 {\it Computers and Intractability: a
 Guide to the theory and NP-completeness} 
(Freeman, San Francisco, 1979); 
C.H. Papadimitriou, {\it Computational Complexity} (Addison-Wesley, 1994).


\bibitem{SP}M\'ezard M, Zecchina R, Phys. Rev. {\bf E 66 } 056126 (2002);
  M\'ezard M, Parisi G, Zecchina R, SCIENCE {\bf 297} 812 (2002)

\bibitem{MZ} Monasson R, Zecchina R, Phys. Rev. {\bf  E 56}  1357 (1997)
  
\bibitem{Mulet}Mulet R, Pagnani A, Weigt M, {\it et al.}, Phys. Rev. Lett.
  {\bf 89 } 268701 (2002)

\bibitem{Anderson} Y. Fu and P. W. Anderson, J. Phys. A {\bf 19}, 1605-1620 (1986). 

\bibitem{MePaVi} 
M M\'ezard, G Parisi, MA Virasoro, World Scientific (1987)
{\em Spin glass theory and beyond}

\bibitem{all} F. Krzakala, A. Montanari, F. Ricci-Tersenghi, G. Semerjian and
  L. Zdeborov\'a, cond-mat/0612365. Proc. Natl. Acad. Sci. 104, 10318 (2007).

\bibitem{LS}
B. D. Lubachevsky and F. H. Stillinger, J. Stat. Phys. {\bf 60},  561 (1990).

\bibitem{SW}Stillinger FH, Weber TA, 
J. Chem. Phys.{\bf 83}  4767 (1985) 

\bibitem{annealed}A-S Sznitman, {\em 
 Brownian motion, obstacles and random media}, Springer, Berlin, (1998).

\bibitem{Moukarzel1} This is perhaps related to the formalism in: Duxbury PM,
  Jacobs DJ, Thorpe MF, {\it et al.}  Phys. Rev. {\bf E 59} 2084 (1999). A
  potential is also defined in: Brito C, Wyart M Europhys. Lett. {\bf 76} 986
  (2006)
  
\bibitem{TheseM} For an overview, see M. Wyart, Annales de Physique, Vol. 30
  No. 3 (2005).

\bibitem{De}C. De Dominicis, M. Gabay, T. Garel and H. Orland,
 J. Phys. (France), {\bf 41}, 922-30 (1980)

\bibitem{BM}
AJ Bray, MA Moore - J. Phys. C Solid St. Phys. {\bf 13} L469
(1980)
 
\bibitem{Cuku2} LF Cugliandolo and J. Kurchan,
J. Phys. {\bf A 27}  5749 (1994)

\bibitem{Leticia}
L. F. Cugliandolo,  {\em
Lecture notes in Slow Relaxation and non equilibrium dynamics in condensed matter},
in  Les Houches Session 77 July 2002, 
J-L Barrat, J Dalibard, J Kurchan, M V Feigel'man eds.
cond-mat/0210312 

\bibitem{walter}         Kob W, Sciortino F, Tartaglia P, 
Europhys. Lett. {\bf  49} 590 (2000) 

\bibitem{Blythe} The literature on the metastable states of the
  Sherrington-Kirkpatrick model covers  a quarter of a
  century.  For recent works , see: A. Crisanti, L. Leuzzi, G. Parisi, T.
  Rizzo Phys. Rev. B 70, 064423 (2004) M. Mueller, L. Leuzzi, A Crisanti,
  Phys. Rev. B 74, 134431 (2006) Cavagna A, Giardina I, Parisi G Phys. Rev.
  Lett. {\bf 92 } 120603 (2004);
 \\ The barrier to escape a state of free energy per spin $f$ 
 scales as $ B ~
(f-f_0)^{-1/3}$, where  $f_0$
is the free energy density of the  lowest state; 
a finite number of flips will unstabilize states with $f>f_0$, see:
  T. Aspelmeier, R. A. Blythe, A. J. Bray, M. A. Moore, 
Phys. Rev. {\bf B 74}, 184411 (2006).  Thus,
  these calculations are in agreement with the conclusions from the dynamics.

\bibitem{foot2}The algorithm is the adaptation of an efficient
 strategy introduced for satisfiability in \cite{Asat}: We 
choose  randomly one of
 the spins that has the same colors that one, or more, of its neighbors and
 change
 its color randomly: the move is accepted with probability one if this improves
the coloring ---if more 
spins are satisfied--- and with a probability $\alpha$ otherwise.
The parameter  $\alpha$ is  tuned for more efficiency,
 see \cite{all-col} for details. 

\bibitem{foot4}In the thermodynamic limit,
it is not at present clear  whether the point reached after an annealing from
 high temperatures to a given low temperature $T<T_d$, followed by a
long wait at $T$
depends on the speed of the annealing. If this were so, then the threshold
needs a process-dependent definition in this model ~\cite{anneal}.

\bibitem{KuPaVi}J. Kurchan, G. Parisi and MA Virasoro,
J. de Physique {\bf  I}  3  1819 (1993)

\bibitem{CrSo}A. Crisanti and H-J Sommers, 
J. de Physique {\bf  I} 5  805 (1995) 

\bibitem{KuLa}
Kurchan J, Laloux L,
J. Phys. {\bf A 29}  1929 (1996) 

\bibitem{Cuku1}LF Cugliandolo and J. Kurchan,
 Phys. Rev. Lett. {\bf 71} 173  (1993) 

\bibitem{foot0}This can always be done by doing Monte Carlo dynamics
with a temperature  adjusted to keep energy constant

\bibitem{Let} L.F. Cugliandolo, D. Grempel, G. Lozano and H. Lozza,
to be published.

\bibitem{ccc} The transition where this happens
along an equilibrium path was computed in \cite{all-col}. Here the idea
is the same but along a nonequilibrium path.

\bibitem{all-col}
L. Zdeborov\'a and F. Krzakala, {\it Phase transitions in the coloring of
  random graphs}, arXiv:0704.1269v1 (2007).

\bibitem{all-sat}
A. Montanari,  F. Ricci-Tersenghi and G. Semerjian, in preparation.

\bibitem{SAAD} J. van Mourik and D. Saad, Phys. Rev. E {\bf 66}, 056120 (2002).

\bibitem{AF} Montanari A, Ricci-Tersenghi F
Eur. Phys. Jour.{\bf  B 33}  339 (2003) 

\bibitem{ALAIN} A. Barrat, {\it The spherical p-spin model}, cond-mat/9701031.


\bibitem{anneal}
Montanari A, Ricci-Tersenghi F
Phys. Rev. {\bf  B 70 } 134406 (2004);
  Capone B, Castellani T, Giardina I, {\it et al.}
Phys. rev.  {\bf B 74 } 144301 (2006) 

\bibitem{Full_sat} Andrea Montanari, Giorgio Parisi, Federico Ricci-Tersenghi
 J. Phys. {\bf A 37}, 2073 (2004).

\bibitem{Full_col}  Florent Krzakala, Andrea Pagnani, Martin Weigt
 Phys. Rev. {\bf E 70}, 046705 (2004).

\bibitem{MMM}
R. Monasson and D. O'Kane,
Europhys. Lett. {\bf 27}  (1994); 
Biroli G, Monasson R and Weigt M (2000), Eur. Phys. J. {\bf B 14}, 551;
M\'ezard M, Palassini M and Rivoire  (2005), 
Phys. Rev. Lett. {\bf 95}, 200202;
and Ref. \cite{all}.


\bibitem{Sri}Sastry S, Debenedetti PG, Stillinger FH Nature {\bf 393} (6685)
  554 (1998)

\bibitem{walksat}B Selman, HA Kautz and B Cohen, proc. of AAAI-94, Seattle (1994);

\bibitem{Asat}Jonh Ardelius and Erik Aurell, Phys. Rev. E 74, 037702 (2006).
  
\bibitem{BPalone} BP alone does not give a solution to the problem,
  but only estimates the probability that a given variable takes a
  given color over the whole set of solutions. Finding the solution
  thus requires a decimation of these variables (see \cite{all-col}
  for details). To find a proper coloring of a graph of $N$ variables,
  we applied the following procedure recursively $N$ times: (i) run BP
  and (ii) fix the most biased variable to its most likely value. We
  have also tried to fix the variables according to their BP
  probability estimate ---and not systematically to its most biased
  value--- but, interestingly, the iterative application of Walk-COL
  is less effective when starting from these solutions.  This is a
  sign that the first procedure gives solutions that belong to larger
  clusters.

\bibitem{MS}Montanari A, Semerjian G Phys. Rev. Lett. {\bf 94} 247201 (2005)
  and J. Stat. Phys. {\bf 124}, 1572 (2006).
  
\bibitem{recolorings} In fact the model considered in \cite{MS} is somewhat
  special. More general and detailes results are obtained for the coloring and
  the satisfiability problem in: G.  Semerjian, {\em On the freezing of
    variables in random constraint problems}, arXiv:0705.2147.
  
\bibitem{crysta} In finite dimensional systems one may minimize the conceptual
  complications related to crystallisation is several ways: {\em i)}
  considering high enough dimensions where it is conjectured that the basin of
  attraction
of amorphous
  packings dominates \cite{high}, {\em ii)} working with a dynamics that
  artificially imposes constant values of parameters of crystallinity (denoted
  $Q$ in \cite{Sal1}), {\em iii)} using a mixture of particles that either does
  not crystallize or if it does its order is so complicated as to be
  dynamically irrelevant.
  
\bibitem{high} Skoge M, Donev A, Stillinger FH, Torquato S Phys. rev. {\bf E
    74} 041127 (2006)
 
\bibitem{hypo} Chaikin PM, Donev A, Man WN, Stillinger FH and Torquato S
  Industrial and engineering chemical research, 45 : 6960 11 (2006); Donev A,
  R. Connely, FH Stillinger and S. Torquato, {\em to be published}.

\bibitem{amorphous} Skoge M, Donev A, Stillinger FH and Torquato S Phys. rev.
  {\bf E 74 } 41127 (2006); and references therein.

\bibitem{Sal1} S. Torquato, T. M. Truskett, and P. G. Debenedetti Phys. Rev.
  Lett.{\bf 84}, 2064 (2000)

\bibitem{debate} Ref. \cite{OHern} was criticised in: Donev A, Torquato S,
  Stillinger FH, {\it et al.}  Phys. Rev. {\bf E 70} 043301 (2004). See the reply:
  C. S. O'Hern, L. E. Silbert, A. J. Liu, and S. R. Nagel Phys. Rev. {\bf E
    70}, 043302 (2004)

\bibitem{LN}Liu AJ, Nagel SR, Nature {\bf 396 } 6706 (1998)

%%%%%%%%%%%%%%%%%%%%%%%%%%%%%%%%%%%%%%%%%%%%%%%%%%%{EvCoMo,GaCo3,FT}

\bibitem{BiMeRi}O. Rivoire, G. Biroli, O. C. Martin, M. M\'ezard Eur. Phys.
  J.{\bf B 37}, 55-78 (2004)
 

 
\bibitem{Wyart} Wyart M, Nagel SR, Witten TA Europhys. Lett {\bf 72} 486
  (2005)
 
\bibitem{Moukarzel} Moukarzel CF Phys. Rev. Lett. {\bf 81} 1634 (1998)

\bibitem{foot1}They obtain what seems an equivalent situation with a different
  `deflating' process, but we shall not discuss that one here.

\bibitem{foot5} Here `annealing' refers to letting a system adapt slowly to
  each new energy, and is not to be confused with the `annealed
  approximation', which consists of optimizing both configuration {\em and}
  problem simultaneously.
 
\bibitem{i-sat} J. Whittemore, J. Kim, K. Sakallah ``SATIRE: a new incremental
  satisfiability engine'' in Proc. of the 38th conference on Design
  automation, pages 542--545, 2001.  N. Eén, N. Sörensson ``Temporal Induction
  by Incremental SAT Solving'' First International Workshop on Bounded Model
  Checking, ENTCS issue 4 volume 89.

\bibitem{ReSi} Cocco S, Monasson R Theoretical Computer Science {\bf 320} 345
  (2004)
 
\bibitem{FrPa}Franz S, Parisi G Physica {\bf A 261} 317 (1998)

\bibitem{PaVi} G. Parisi and MA Viarsoro, Journal de Physique (Paris) {\bf 50}
  3317 (1989)

\bibitem{foo} One may then ask if an infinite temperature inherent structure 
of, say, a Lennard-Jones system, should also be critical. The studies of
 vibration modes of inherent structures of  
temperature $T$ seem to show an excess of very low frequency modes at higher 
starting temperatures $T$ 
(see Fig. 3 of Ref.~\cite{walter}), but this requires a more detailed 
study. 

\bibitem{Remi}R Monasson
Phys. Rev. Lett. {\bf 75}  2847 (1995) 

\bibitem{perco} J. M. Schwarz, A. J. Liu, L. Q. Chayes. cond-mat/0410595.
 
\bibitem{FEDE} D. Achlioptas and F. Ricci-Tersenghi, CC/0611052.
  
\bibitem{uff} The clustering point is the celebrated Mode-Coupling (dynamic)
  transition point of glass theory. This has been recently extended to dilute
  systems in: Andrea Montanari, Guilhem Semerjian, J. Stat. Phys. {\bf 125},
  23 (2006).

\bibitem{Zampo} An additional puzzle is brought about by the replica
  calculations, where states are isostatic but not critical, see:
  Parisi G, Zamponi F, J. Stat. Mech 3017 (2006); J. Chem. Phys. 123
  144501 (2005).


\bibitem{Zampouno} See the discussion for hard spheres in  
F.Zamponi, Phys.Rev.E 75, 043101 (2007).



\end{thebibliography}
\end{document}